# Antisymmetric interlayer exchange coupling spontaneously built in synthetic antiferromagnetic structure by film growth

Takeshi Seki[1,2,a)], Hiroto Masuda[3], Varun K. Kushwaha[1], and Takumi Yamazaki[1]

[1]*Institute for Materials Research, Tohoku University, Sendai 980-8577, Japan*

[2]*Center for Science and Innovation in Spintronics, Tohoku University, Sendai 980-8577, Japan*

[3]*Department of Materials Physics, Nagoya University, Nagoya 464-8603, Japan*

[a)] *takeshi.seki@tohoku.ac.jp*

**Abstract**

The antisymmetric-type long-range exchange interaction between two ferromagnetic layers through a nonmagnetic layer, called antisymmetric interlayer exchange coupling (AIEC), has recently been discovered and attracted much attention. This paper reports that the AIEC is naturally built in synthetic antiferromagnets depending on the thin film growth condition. For the synthetic antiferromagnets consisting of Pt/Co/Ir/Co/Pt layers, two kinds of film growth parameters are examined: the effects of epitaxial growth and oblique incident of sputter-deposition. The present results indicate that the spatial fluctuations in thicknesses are one of the major sources inducing the AIEC while the epitaxial growth, which leads to the sharp interfaces and the uniformity in layer thicknesses, effectively suppresses the unexpected AIEC. In addition, the oblique sputter-deposition provides the large AIEC and the anisotropic distribution of AIEC direction in the case of textured films. The findings in this study provide key factors for designing the AIEC and is useful to develop emerging computing technologies with artificial three-dimensional topological magnetic structures.

**Main text**

I.  Introduction

Long-range exchange interaction between ferromagnetic layers separated by a nonmagnetic layer, which is called the interlayer exchange coupling (IEC), leads to the antiparallel alignment of magnetization vectors for the adjacent ferromagnetic layers [1-4]. The artificial antiferromagnetic spin structure is well known as a synthetic antiferromagnet (SyAF), and is a vital component for the contemporary spintronic applications such as magnetic random access memories because the antiparallel alignment of magnetization vectors reduces the stray magnetic field from the memory cell and at the same time makes the memory cell insensitive to external magnetic field disturbance. In addition to the practical applications, the SyAFs have recently attracted renewed interest as an academic subject in the research field of antiferromagnetic spintronics. Antiferromagnetic spintronics [5,6] exploits the characteristics of antiferromagnet: low magnetic susceptibility, lack of stray magnetic field, and high frequency dynamics coming from the antiferromagnetic resonance. These characteristics will allow us to develop next-generation spintronic applications with higher storage density and faster operation speed than the conventional ferromagnet-based spintronic applications. Although most works were done for bulk antiferromagnets such as CuMnAs [7,8] and NiO [9,10], considering that one carries out the systematic investigation, *e.g.*, interplay between the antiferromagnetic spin structure and the conduction electron spin, a SyAF is favorable because the magnetic anisotropy and exchange coupling constants for the SyAF can be controlled by tuning the layer thicknesses and the well-defined magnetization alignments and domain structures can be exploited [11-15].

The IEC was discovered by Peter Grünberg and co-workers in 1986 for the ferromagnetic Fe layers separated by the non-ferromagnetic Cr layers [1]. The exchange energy of IEC is given by $E = -J(\mathbf{m}_1 \cdot \mathbf{m}_2)$, where $\mathbf{m}_1$ and $\mathbf{m}_2$ represent the magnetization vectors of two FM layers. The sign of the exchange coupling constant $J$ determines whether $\mathbf{m}_1$ and $\mathbf{m}_2$ are coupled ferromagnetically or antiferromagnetically. After the discovery of IEC, many researchers have studied the IEC in various combinations of ferromagnets and nonmagnets [3,4]. The recent studies on antiferromagnetic spintronics discovered the antisymmetric type of long-range exchange interaction in the SyAFs, called antisymmetric IEC (AIEC) [16-18]. The AIEC emerges in the SyAFs with in-plane spatial inversion symmetry breaking. The AIEC energy is expressed as $\mathbf{D}_{AIEC} \cdot (\mathbf{m}_1 \times \mathbf{m}_2)$, where $\mathbf{D}_{AIEC}$ is the AIEC vector determined by system symmetry. In the Cartesian coordinate system, given that $\mathbf{m}_1$ and $\mathbf{m}_2$ pointing in the $z$ direction are coupled via the IEC and the inversion symmetry is broken along the $y$ direction, one may anticipate the existence of $\mathbf{D}_{AIEC}$ along the $x$ direction. As a result, the noncolinear magnetization alignment between $\mathbf{m}_1$ and $\mathbf{m}_2$ appears.

Chiral magnetic structures due to the AIEC will be a building block of artificial three-dimensional topological magnetic structures and have potential use for emerging computing technologies [19-21]. For some spintronic applications utilizing only the collinear magnetization alignments, on the other hand, such a chiral magnetic structure is an obstacle for stable device operation. These points suggest the significance of controllability of AIEC in the SyAF. Although there are several reports on the observation of AIEC [22-32] and its application to the magnetization switching technique based on the spin-orbit interaction, that is spin-orbit torque magnetization switching [24,25,28,29,31,32], the origin of in-plane spatial inversion symmetry breaking has not been obvious for most studies and the control of $\mathbf{D}_{AIEC}$ has not fully been achieved. Our previous work [23] utilized the wedge-shaped ferromagnetic Co and nonmagnetic Ir layers in the perpendicularly magnetized Pt/Co/Ir/Co/Pt structures, and expected that the wedge shapes provide the in-plane spatial inversion symmetry breaking along the slope directions. Although we concluded that the insertion of wedge-shaped layers is effective to enhance the magnitude of AIEC, the $\mathbf{D}_{\mathrm{AIEC}}$ showed the different direction for that expected from the slope directions. These experimental facts imply that there are other sources inducing the finite AIEC. This idea was also supported by the fact of the small but non-negligible AIEC even for the non-wedged samples [23]. Another research group also has pointed out the difficulty in controlling the direction of $\mathbf{D}_{\mathrm{AIEC}}$ [22]. Thus, the clarification of source for the AIEC without a wedge-shaped layer insertion is inevitable for improving the controllability of AIEC and further enhancing the magnitude of AIEC.

This paper reports the experimental investigation of the AIEC naturally built in Pt/Co/Ir/Co/Pt structures by thin film growth. Two kinds of thin film growth parameters are examined. One is whether the layers are epitaxially grown or not. The Pt/Co/Ir/Co/Pt layers were grown on different substrates: an $Al_2O_3$ (0001) single crystal substrate and a thermally-oxidized Si (Si-O) substrate. The other is the effect of oblique incidence of sputtered atoms to the substrate. The thin film growth was carried out with or without the substrate rotation. The magnitude and direction of AIEC are investigated using four kinds of samples, and the contribution of thin film growth condition to the AIEC is discussed. Based on the key factors for enhancing or suppressing the AIEC, large AIEC is demonstrated.

## II.  Experimental Procedure

As schematically illustrated in **Fig. 1(a)**, the common layer stacking among the samples is Pt (4.5)/Co ($t_{\mathrm{Co}}$)/Ir ($t_{\mathrm{Ir}}$)/Co (0.5)/Pt (2.7) (in nanometer). The thin films were prepared on $Al_2O_3$ (0001) single crystal substrates or Si-O substrates using a magnetron sputtering system with the base pressure of $\sim 2 \times 10^{-5}$ Pa. The magnetron sputtering system is equipped with three sputtering cathodes. The cathode #1 has a revolving target changer with Ta and Pt sputtering targets, and Ta or Pt is sputter-deposited at the incident angle of 0º, that is normal to the substrate surface. The cathode #2 and the cathode #3 have Co and Ir sputtering targets, respectively, and the incident angles are set to be 45º as illustrated in **Fig. 1(b)**.

For the samples on the Si-O substrates, which are labelled Sample A, the 2 nm-thick Ta buffer layer was deposited before the deposition of 4.5 nm-thick bottom Pt layer. All the layers were deposited at room temperature (RT). In contrast to Sample A, the samples on the $Al_2O_3$ (0001) single crystal substrates, which are labelled Sample B, do not involve the Ta buffer layer. Instead of insertion of Ta buffer layer, the 4.5 nm-thick bottom Pt layer was deposited at 300ºC, and the other layers were deposited at RT.

In order to examine the effect of oblique incidence of sputtered atoms to the substrate, Sample A and Sample B were deposited with or without the substrate rotation. For the samples deposited with the substrate rotation, which are called Sample A-w and Sample B-w, we expect that the effect of oblique incidence is averaged out and disappears. On the other hand, for the samples deposited without the substrate rotation, the substrate rotation was stopped only during the deposition of Co layers and the side of substrate was aligned to the incident direction of sputtered atoms as shown in **Fig. 1(b)**. These samples without the substrate rotation are labelled Sample A-w/o and Sample B-w/o. Since one may think the oblique incidence gives rise to the gradient of layer thickness, prior to the thin film preparation, the position dependence of film thickness was measured for the sample deposited without the substrate rotation on the Si-O substrate with the size of 10 mm × 10 mm. The thickness variation as a function of position in the 10 mm × 10 mm substrate is plotted in **Fig. 1d**. The oblique incidence does not induce the clear thickness gradient.

The thin films were patterned into a Hall-bar shape using photolithography and Ar-ion milling. In this study, the devices with different channel widths $d$ were fabricated, where $d$ = 5, 10 and 20 μm. **Figure 1(c)** displays the optical image of the device with $d$ = 10 μm together with the terminal configuration for the Hall effect measurement. The electrical contact pads of Cr (20 nm)/Au (200 nm) were fabricated using photolithography and ion-beam sputtering. The longitudinal direction of the channel was designed to coincide with the in-plane projection component of the oblique incidence. In order to evaluate the AIEC, the additional in-plane magnetic field $H_{ip}$ was applied under the sweep of out-of-plane magnetic field $H_z$ [18,23]. The definition of in-plane magnetic field angle $\phi$ is also shown in **Fig. 1(c)**. $\phi = 0°$ corresponds to the longitudinal direction of the channel, that is the same direction as the in-plane projection component of the oblique incidence.

The structural characterization was carried out using reflection high-energy electron diffraction (RHEED) and scanning transmission electron microscopy (STEM) with energy dispersive x-ray spectroscopy (EDS). Magnetic hysteresis loops were measured using the polar magneto-optical Kerr effect (MOKE). The Hall effect was measured employing the physical properties measurement system with the maximum field of 9 T, and the probe station with a magnet which produces a magnetic field in any direction. All the measurements were performed at RT.

### III.   Results and Discussion

#### 1. Structural Characterization

**Figure 2** displays the RHEED patterns for the bottom Pt, the bottom Co, the Ir and the top Co layers of **(a)** Sample A on Si-O and **(b)** Sample B on Al$_2$O$_3$ (0001). $t_{Co}$ was 0.9 nm for both samples whereas $t_{Ir}$ was 0.5 nm for Sample A and 0.7 nm for Sample B. Sample A exhibits diffusive streaks overlapped with thin arcs. During the RHEED observation, these patterns do not vary with changing the direction of observation (incident angle of electron beam), which indicates that Sample A has a texture structure. On the other hand, Sample B shows the clear streaks for all the layers than those for Sample A, and the patterns vary by changing the direction of observation. These results indicate that Sample B was epitaxially grown on the $Al_2O_3$ (0001) single crystal substrate. Compared with the previous structural analysis for the Co/Ir/Co stacks grown on the $Al_2O_3$ (0001) [33], Sample A is the (111)-oriented texture film and Sample B is the (111)-oriented epitaxial film.

The cross-sectional high-angle annular dark-field (HAADF)-STEM images and STEM-EDS elemental maps for Sample A and Sample B are given in **Fig. 3**, where $t_{Co}$ and $t_{Ir}$ were 0.9 nm and 0.7 nm, respectively, for both samples. These samples were deposited with the substrate rotation. One can see the clear layered structures for both Sample A and Sample B. Particularly, even 0.5 nm-thick Co layers form the continuous morphologies. Remarkable differences between these samples are that the well-ordered atomic arrangement is observed for Sample B thanks to the epitaxial growth and that Sample B possesses the sharper interfaces than Sample A. It is noted that there are non-negligible spatial fluctuations in thicknesses for both Sample A and Sample B.

## 2. Hall loops and magneto-optical Kerr effect loops

**Figure 4** shows the transverse resistance $R_{xy}$ as a function of $H_z$ for **(a)** Sample A-w, **(b)** Sample A-w/o, **(c)** Sample B-w, and **(d)** Sample B-w/o, where $t_{Co}$ and $t_{Ir}$ are 1.3 nm and 0.5 nm, respectively, for all the samples. These Hall loops were measured with the magnetic field applied in the range of -9 T $\leq \mu_0 H_z \leq$ 9 T. The additional $H_{ip}$ was not applied. The insets of **Fig. 4** are the polar MOKE loops measured in the magnetic field range of -2 T $\leq \mu_0 H_z \leq$ 2 T. From the comparison between the Hall loops and the polar MOKE loops, it is confirmed that the anomalous Hall effect (AHE) proportional to the perpendicular component of net magnetization of two Co layers is the dominant in $R_{xy}$. Also, both Hall loops and polar MOKE loops indicate that the two Co magnetizations are antiferromagnetically aligned for all the samples at low $H_z$. Since $t_{Ir}$ = 0.5 nm is the first peak of $J$ for the Co/Ir/Co systems and the strong antiferromagnetic coupling is achieved [14,33], even the application of $\mu_0 H_z = 9$ T is not sufficient to align the Co magnetization vectors in parallel. Thus, the Hall loops for all the samples are not saturated in the field range of -9 T ≤ $\mu_0 H_z$ ≤ 9 T. Since the similar Hall loops were obtained for all four samples, all four samples possess the comparable strength of antiferromagnetic coupling.

### 3. Evaluation of antisymmetric interlayer exchange coupling

The magnitude of AIEC is evaluated by measuring the $R_{xy}$ - $H_z$ curves with the application of additional $H_{ip}$. As explained in **Fig. 1(c)**, $\phi = 0°$ is aligned with the longitudinal direction of the channel, corresponding to the in-plane projection component of the oblique incidence in the cases of Sample A-w/o and Sample B-w/o. **Figure 5(a)** shows the $R_{xy}$ - $H_z$ curves with $\mu_0 H_{ip} = 50$ mT applied at $\phi = 150°$ and $330°$ for the Hall device of Sample A-w/o with $d = 20$ μm. $t_{Co}$ and $t_{Ir}$ are 1.3 nm and 0.5 nm, respectively. Under the additional $H_{ip}$ application, the presence of AIEC leads to the asymmetry in the up-to-down (UD) and down-to-up (DU) switching fields for the net magnetization because the preferred magnetic chirality is dictated by the AIEC. According to the previous work [18,23], the application of $H_{ip}$ lifts the degeneracy in the energies of magnetization switching for tail-to-tail and head-to-head configurations of two Co layers, resulting in the asymmetry in the UD and DU switching events. The magnitude of unidirectional shift of $R_{xy}$ - $H_z$ curve is related to the magnitude of AIEC. The UD and DU switching fields $H_{sw}$ are obviously different, and the directions of shift in the $R_{xy}$ - $H_z$ curves are also different between the cases at $\phi = 150°$ and $330°$. From the measurement of $R_{xy}$ - $H_z$ curve under the $H_{ip}$ application, the values of $\mu_0 H_{sw}$ as a function of $\phi$ are plotted for the UD and DU switching events in **Fig. 5(b)** and **Fig. 5(c)**, respectively. The experimental data are numerically fitted with $\mu_0 H_{sw} = \mu_0 H_{sw}^0 + \mu_0 H_{shift} \cos(\phi - \phi_{AS})$. $\mu_0 H_{shift}$ represents the magnitude of AIEC, and $\phi_{AS}$ is the direction of AIEC, which corresponds to the in-plane direction orthogonal to $\mathbf{D}_{AIEC}$. $\mu_0 H_{sw}^0$ is the baseline in the angular dependence of switching field. $\mu_0 H_{shift}$ = 15.4 mT (14.7 mT) and $\phi_{AS}$ = 151° (153°) are evaluated for the UD (DU) switching events for the Hall device of Sample A-w/o with $d$ = 20 μm.

**Figure 6** summarizes the values of $\mu_0 H_{shift}$ and $\phi_{AS}$ plotted as pole figures for the Hall devices of **(a)** Sample A-w, **(b)** Sample A-w/o, **(c)** Sample B-w, and **(d)** Sample B-w/o, where $t_{Co}$ and $t_{Ir}$ are 1.3 nm and 0.5 nm, respectively. In order to discuss the statistical trends of $H_{shift}$ and $\phi_{AS}$, many devices fabricated on the identical substrate were measured: 25 devices for Sample A-w, 23 devices for Sample A-w/o, 36 devices for Sample B-w, and 35 devices for Sample B-w/o. Those data points involve both the UD and DU switching events for the Hall devices with $d$ = 5 μm (circles), 10 μm (triangles), and 20 μm (squares).

## 4. Discussion

First let us discuss the effect of epitaxial growth on the AIEC from the results for Sample A-w and Sample B-w, which are the (111)-oriented texture film on the Si-O substrate and the (111)-oriented epitaxial film on the $Al_2O_3$ (0001) substrate, respectively. Both samples exhibit non-negligible $H_{shift}$, and the values of $\mu_0 H_{shift}$ are randomly distributed with respect to $\phi_{AS}$. The trend of random distribution is obvious for Sample B-w. In the case of Sample B-w, the maximum $\mu_0 H_{shift}$ is 8.5 mT and most data points are less than 5 mT. In addition, no particular $\phi_{AS}$ with large AIEC is observed. Compared to Sample B-w, the result for Sample A-w has more scattered data points. Although many devices exhibit $\mu_0 H_{shift}$ less than 5 mT, one device exhibits a large $\mu_0 H_{shift}$ close to the 15 mT. Since the films of Sample A-w and Sample B-w were deposited with the substrate rotation, it is natural that any anisotropic properties induced by the film growth process, *e.g.*, effect of oblique incident or induced magnetic anisotropy, are averaged out, and one may anticipate that no AIEC exists if an ideally uniform layer stacking is achieved. However, the present samples are not the case. These experimental results indicate that there exists an unexpected AIEC. One useful finding in this study is that the epitaxial growth is effective to suppress the unexpected AIEC. The HAADF-STEM images shown in **Fig. 3** suggest that both Sample A-w and Sample B-w involve the spatial fluctuations in thicknesses, and that the fluctuation is more remarkable for Sample A-w. Considering these experimental facts, the thickness fluctuation is one of the possible sources of breaking the in-plane spatial inversion symmetry, leading to the non-negligible AIEC. Compared with the textured film, the epitaxial growth leads to the sharp interfaces and the uniform layer thicknesses, resulting in the smaller magnitude of AIEC for the epitaxial film.

Next, in order to examine the effect of oblique incidence of sputtered atoms to the substrate, we compare the results between Sample A-w and Sample A-w/o. The remarkable anisotropic distribution in $H_{shift}$ is observed for Sample A-w/o, which was the textured film deposited without the substrate rotation. The largest $\mu_0 H_{shift}$ among all four samples is obtained for Sample A-w/o, which is 16.7 mT. Several devices for Sample A-w/o also exhibit the large $\mu_0 H_{shift}$ close to 15 mT. Although $\phi_{AS}$ for the devices showing large $\mu_0 H_{shift}$ does not strictly follow the direction of in-plane projection component of the oblique incidence, which is $\phi_{AS} = 0º$ or $180º$, the anisotropic distribution in $H_{shift}$ mainly centered around $\phi_{AS} = 180º$ and the large values of $\mu_0 H_{shift}$ indicate that the oblique sputter-deposition obviously contributes to the induction of AIEC.

A coupled of previous studies also reported the growth-induced AIEC for CoFeB / Pt / Co [Ref. 24] and [Pt / Co]$_3$ / Ru / [Co / Pt]$_2$ [Ref. 31], in which the CoFeB layer and Pt layers, respectively, were grown by the oblique sputter-deposition. Our experimental results for the textured film (Sample A-w/o) are qualitatively consistent with the tendency of the previous studies [24,31]. However, the control of $\phi_{AS}$ for our case is yet to be enough compared to the previous work [24], which reported the well-controlled $\phi_{AS}$ by oblique deposition. Although the reason for the difference between the present work and the previous ones is not clear at present, since the effect of oblique sputter-deposition is largely affected by the specifications of equipment, *e.g.*, the pressure of process gas (Ar gas in the present study) during deposition, the distance between the substrate and the sputtering target, those technical differences in the sputter-deposition condition might be factors for difference in the controllability of $\phi_{AS}$. In contrast to the control of $\phi_{AS}$, the Sample A-w/o exhibit the larger AIEC than the previous study [31]. According to Ref. 31, the interlayer Dzyaloshinskii–Moriya interaction (DMI) effective field $H_{DMI}$, which corresponds to the effective field of AIEC, is evaluated from the equation of $(\mu_0 H_{sw}^{UD} + \mu_0 H_{sw}^{DU})/2 = \mu_0 H_{DMI} \cos(\phi - \phi_{AS})$, where $\mu_0 H_{sw}^{UD}$ and $\mu_0 H_{sw}^{DU}$ are the values of $\mu_0 H_{sw}$ for the UD and DU switching events, respectively. For the device of Sample A-w/o shown in **Fig. 5**, $\mu_0 H_{DMI}$ is estimated to be 15 mT, which is larger than the previous values [31]. This large $\mu_0 H_{DMI}$ is attributable to the large IEC (large saturation field more than 9 T shown in **Fig. 4**) thanks to the Ir spacer layer because there exists the correlation between IEC and AIEC [23,26,27].

Another important finding is obtained from the result of Sample B-w/o, which is the epitaxial film deposited without the substrate rotation. Although the distribution of $H_{shift}$ for Sample B-w/o is anisotropic in comparison to Sample B-w, the increase in $H_{shift}$ is not significant in comparison to the cases of the textured films, *i.e.*, Sample A-w and Sample A-w/o. This implies that the oblique sputter-deposition is not so effective in the case of epitaxial films. This is attributable to the well-ordered atomic arrangement of the epitaxial films, and the epitaxial growth may hinder the preferential crystal growth due to the oblique incidence of sputtered atoms onto the substrate.

One may consider that the existence of edges in the microfabricated Hall devices also affects the evaluation of $H_{shift}$ because edges may be possible nucleation sites for reversed magnetic domain. In **Fig. 6(b)** for Sample A-w/o, which possesses the large AIEC, the Hall devices with the widest channel of $d = 20$ μm exhibit the most remarkable anisotropic distribution in $H_{shift}$. This experimental fact implies that a Hall device with a wider channel is suitable to clearly observe the AIEC because of suppression of edges effect on the magnetization reversal process. In other words, the optimization of device design may lead to the improvement of the controllability of $\phi_{AS}$ and the further enhancement of AIEC.

The present experimental results and the above discussion allow us to mention key factors for enhancing the AIEC: (i) non-epitaxial growth, (ii) oblique sputter-deposition, and (iii) large IEC. These also provide useful knowledge when one considers applications. As mentioned in the introductory paragraph, the AIEC is available to form chiral magnetic structures and is expected to have potential use for emerging computing technologies. Toward this goal, a textured film structure and the usage of oblique sputtering enable us to induce the large AIEC in SyAFs. On the other hand, for some spintronic applications with the collinear magnetization alignment, an unexpected chiral magnetic structure should be got rid of for the stable device operation. From the viewpoint of such demand, the epitaxial growth of thin films is an effective way to suppress the AIEC.

## IV. Summary

We carried out the experimental investigation on the AIEC naturally built in the SyAFs consisting of Pt/Co/Ir/Co/Pt layers by thin film growth. By comparing the results for the textured samples and the epitaxial samples, we conclude that the spatial fluctuations in thicknesses are one of the major sources inducing the AIEC. It was also found that the epitaxial growth is effective in suppressing the unexpected AIEC because of sharp interfaces and the uniform layer thicknesses. On the other hand, the oblique sputter-deposition provides the large AIEC and the anisotropic distribution of AIEC direction in the case of textured films. These findings are useful not only to develop emerging computing technologies with artificial three-dimensional topological magnetic structures but also to improve the stable operation of conventional spintronic applications.

**Acknowledgments**

The authors thank R. Umetsu for her help in carrying out a part of the transport measurement, T. Sasaki and T. Endo for their help in doing the film deposition by ion beam sputtering, and S. Ito for his help in observing the scanning transmission electron microscopy images. The device fabrication was partly carried out at the Cooperative Research and Development Center for Advanced Materials, IMR, Tohoku University. This work was supported by JSPS KAKENHI Grant-in-Aid for Scientific Research (A) (JP23H00232).

**References**

[1] P. Grünberg, R. Schreiber, Y. Pang, M. B. Brodsky, and H. Sowers, Phys. Rev. Lett. **57**, 2442 (1986).

[2] P. Bruno, and C. Chappert, Phys. Rev. Lett. **67**, 1602 (1991).

[3] S. S. P. Parkin, Phys. Rev. Lett. **67**, 3598 (1991).

[4] D. E. Bürgler, P. Grünberg, S. O. Demokritov, and M. T. Johnson, "Interlayer Exchange Coupling in Layered Magnetic Structures" in Handbook of Magnetic Materials, volume 13 (ed. K. H. J. Buschow, Elsevier 2001).

[5] V. Baltz, A. Manchon, M. Tsoi, T. Moriyama, T. Ono, and Y. Tserkovnyak, Rev. Mod. Phys. **90**, 015005 (2018).

[6] A. Hirohata, D. C. Lloyd, T. Kubota, T. Seki, K. Takanashi, H. Sukegawa, Z. Wen, S. Mitani, and H. Koizumi, IEEE Access **11**, 117443 (2023).

[7] P. Wadley, B. Howells, J. Železný, C. Andrews, V. Hills, R. P. Campion, V. Novák, K. Olejník, F. Maccherozzi, S. S. Dhesi, S. Y. Martin, T. Wagner, J. Wunderlich, F. Freimuth, Y. Mokrousov, J. Kuneš, J. S. Chauhan, M. J. Grzybowski, A. W. Rushforth, K. Edmond, B. L. Gallagher, and T. Jungwirth, Science **351**, 587 (2016).

[8] S. Y. Bodnar, L. Šmejkal, I. Turek, T. Jungwirth, O. Gomonay, J. Sinova, A. A. Sapozhnik, H. J. Elmers, M. Kläui, and M. Jourdan, Nat. Commun. **9**, 348 (2018).

[9] T. Moriyama, K. Oda, T. Ohkochi, M. Kimata, and T. Ono, Sci. Rep. **8**, 14167 (2018).

[10] T. Yamazaki, T. Seki, T. Kubota, and K. Takanashi, Appl. Phys. Exp. **16**, 083003 (2023).

[11] T. Seki, H. Tomita, T. Shinjo, and Y. Suzuki, Appl. Phys. Lett. **97**, 162508 (2010).

[12] S.-H.Yang, K.-S. Ryu, and S. S. P. Parkin, Nat. Nanotech. **10**, 221 (2015).

[13] R. A. Duine, K.-J. Lee, S. S. P. Parkin, and M. D. Stiles, Nat. Phys. **14**, 217 (2018).

[14] H. Masuda, Y. Yamane, T. Seki, K. Raab, T. Dohi, R. Modak, K. Uchida, J. Ieda, M. Kläui, and K. Takanashi, Appl. Phys. Lett. **122**, 162402 (2023).

[15] J. Godinho, P. K. Rout, R. Salikhov, O. Hellwig, Z. Šobáň, R. M. Otxoa, K. Olejník, T. Jungwirth, and J. Wunderlich, npj Spintronics **2**, 39 (2024).

[16] E. Y. Vedmedenko, P. Riego, J. A. Arregi, and A. Berger, Phys. Rev. Lett. **122**, 257202 (2019).

[17] A. Fernández-Pacheco, E. Vedmedenko, D. Petit, and R. P. Cowburn, Nat. Mater. **18**, 679 (2019).

[18] D.-S. Han, K. Lee, J.-P. Hanke, Y. Mokrousov, K.-W. Kim, W. Yoo, Y. L. W. van Hees, T.-W. Kim, R. Lavrijsen, C.-Y. You, H. J. M. Swagten, M.-H. Jung, and M. Kläui, Nat. Mater. **18**, 703 (2019).

[19] Z. Luo, T. P. Dao, A. Hrabec, J. Vijayakumar, A. Kleibert, M. Baumgartner, E. Kirk, Jizhai Cui, T. Savchenko, G. Krishnaswamy, L. J. Heyderman, and P. Gambardella, Science **363**, 1435 (2019).

[20] A. Fernández-Pacheco, R. Streubel, O. Fruchart, R. Hertel, P. Fischer, R. P Cowburn, Nat. Commun. **8,** 15756 (2017).

[21] A. Hrabec, Z. Luo, L. J. Heyderman, P. Gambardella, Appl. Phys. Lett. **117**, 130503 (2020).

[22] C. O. Avci, C.-H. Lambert, G. Sala, and P. Gambardella, Phys. Rev. Lett. **127**, 167202 (2021).

[23] H. Masuda, T. Seki, Y. Yamane, R. Modak, K. Uchida, J. Ieda, Y. C. Lau, S. Fukami, and K. Takanashi, Phys. Rev. Appl. **17**, 054036 (2022).

[24] Y.-H. Huang, C.-C. Huang, W.-B. Liao, T.-Y. Chen, and C.-F. Pai, Phys. Rev. Appl. **18**, 034046 (2022).

[25] Y.-C. Li, Y.-H. Huang, C.-C. Huang, Y.-T. Liu, and C.-F. Pai, Phys. Rev. Appl. **20**, 024032 (2023).

[26] F. S. Gao, S. Q. Liu, R. Zhang, J. H. Xia, W. Q. He, X. H. Li, X. M. Luo, C. H. Wan, G. Q. Yu, G. Su, and X. F. Han, Appl. Phys. Lett. **123**, 192401 (2023).

[27] S. Liang, R. Chen, Q. Cui, Y. Zhou, F. Pan, H. Yang, and C. Song, Nano Lett. **23**, 8690 (2023).

[28] K. Wang, L. Qian, S.-C. Ying, and G. Xiao, Commun. Phys. **4**, 10 (2020).

[29] W. He, C. Wan, C. Zheng, Y. Wang, X. Wang, T. Ma, Y. Wang, C. Guo, X. Luo, M. E. Stebliy, G. Yu, Y. Liu, A. V. Ognev, A. S. Samardak, and X. Han, Nano Lett. **22**, 6857 (2022).

[30] F. Kammerbauer, W.-Y. Choi, F. Freimuth, K. Lee, R. Frömter, D.-S. Han, R. Lavrijsen, H. J. M. Swagten, Y. Mokrousov, and M. Kläui, Nano Lett. **23**, 7070 (2023).

[31] Y.-H. Huang, J.-H. Han, W.-B. Liao, C.-Y. Hu, Y.-T. Liu, and C.-F. Pai, Nano Lett. **24**, 649 (2024).

[32] X. Zhao, H. Sun, R. Han, H. Qin, L. Wen, H. Wang, D. Wei, and J. Zhao, APL Mater. **12**, 041103 (2024).

[33] H. Masuda, T. Seki, Y.-C. Lau, T. Kubota, and K. Takanashi, Phys. Rev. B **101**, 24413 (2020).

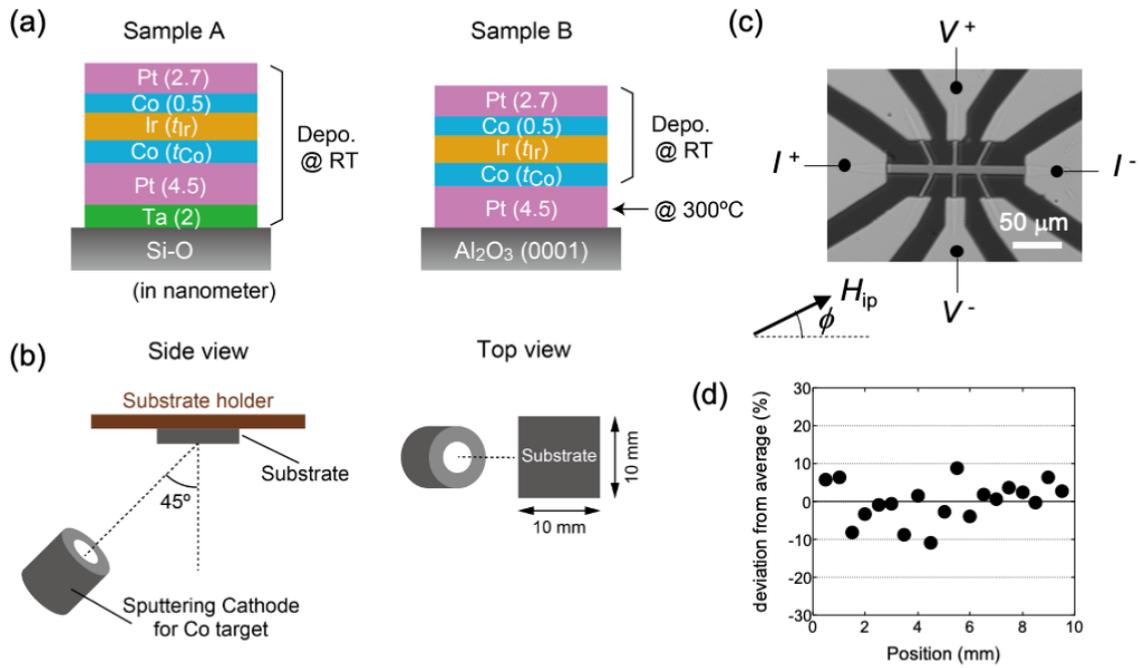

**Figure 1** Schematic illustrations of (a) layer stackings on Si-O substrate and Al$_2$O$_3$ (0001) single crystal substrate, and (b) configuration of substrate and sputtering cathode for Co target. (c) Optical image of the device together with the terminal configuration for the Hall effect measurement. The definition of in-plane magnetic field angle $\phi$ is also shown. (d) The thickness variation as a function of position in the 10 mm × 10 mm substrate.

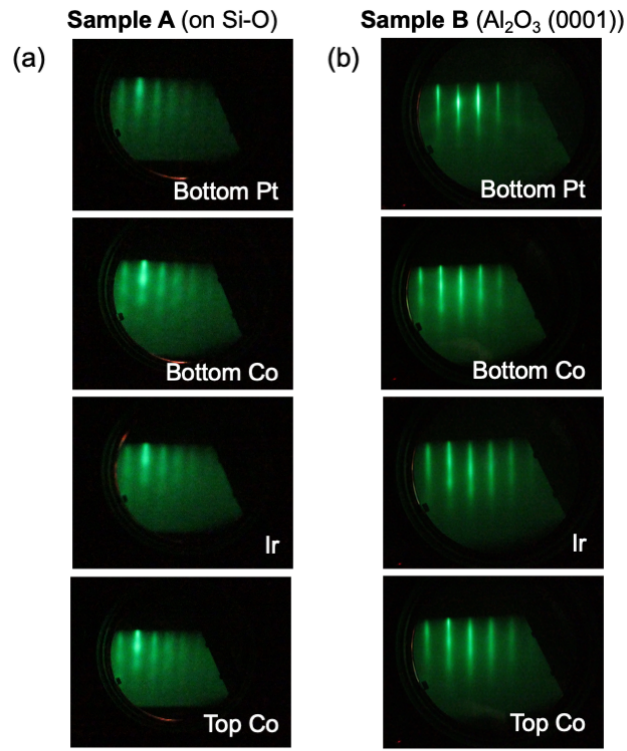

**Figure 2** (a) Reflection high-energy electron diffraction (RHEED) patterns for the bottom Pt, the bottom Co, the Ir and the top Co layers of (a) Sample A on Si-O and (b) Sample B on Al$_2$O$_3$ (0001). $t_{Co}$ was 0.9 nm for both samples, and $t_{Ir}$ was 0.5 nm for Sample A and 0.7 nm for Sample B.

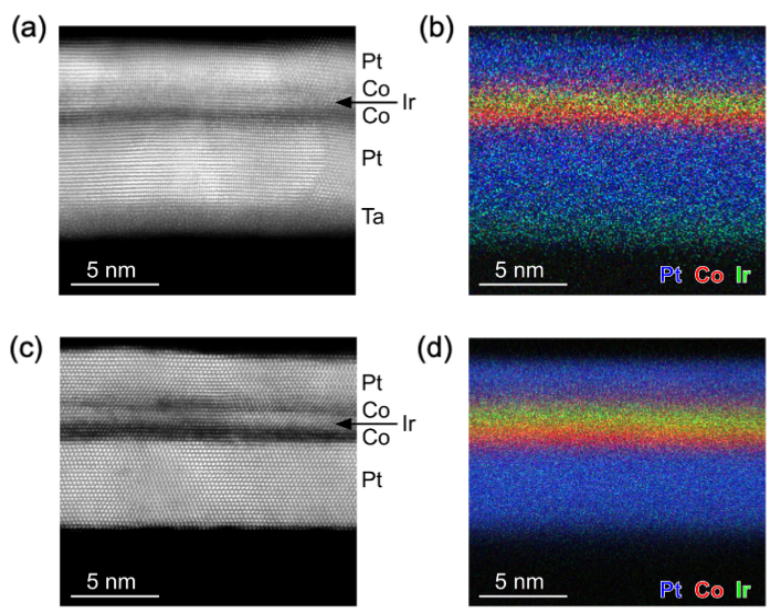

**Figure 3** Cross-sectional high-angle annular dark-field scanning transmission electron microscopy (HAADF-STEM) images and STEM-energy-dispersive x-ray spectroscopy (EDS) elemental maps for Sample A on Si-O and [(a) and (b)] and Sample B on Al$_2$O$_3$ (0001) [(c) and (d)].

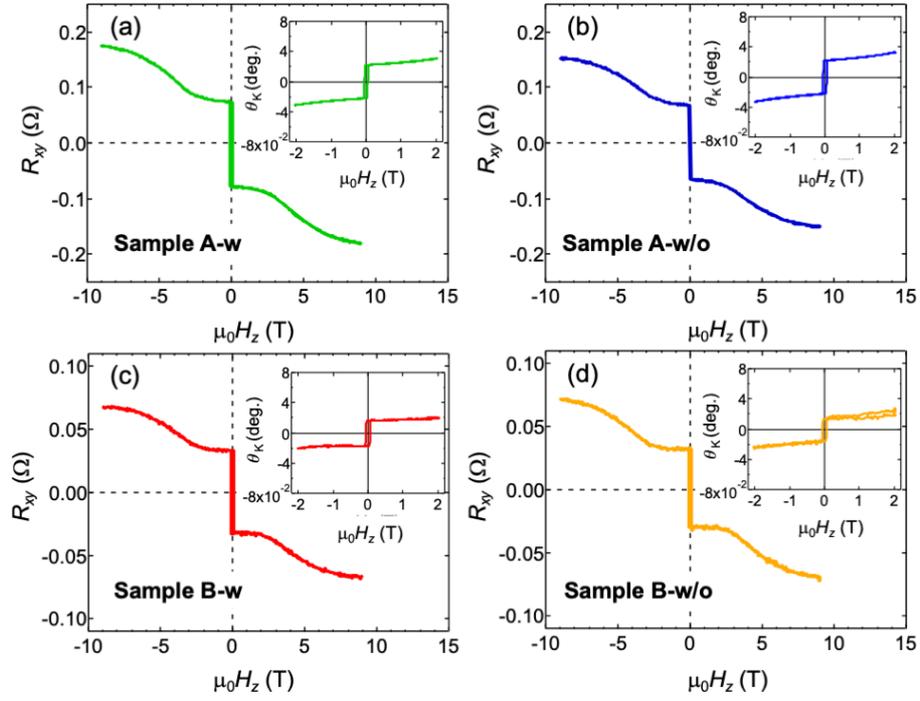

**Figure 4** Transverse resistance $R_{xy}$ as a function of $H_z$ for (a) Sample A-w, (b) Sample A-w/o, (c) Sample B-w, and (d) Sample B-w/o, where $t_{Co}$ and $t_{Ir}$ are 1.3 nm and 0.5 nm, respectively. The additional $H_{ip}$ was not applied. The insets are the polar MOKE loops.

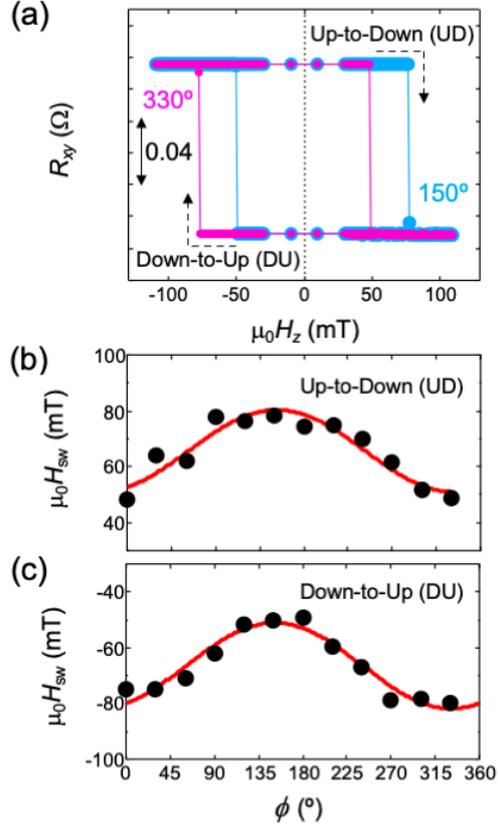

**Figure 5** (a) $R_{xy}$ - $H_z$ curves with $\mu_0 H_{ip}$ = 50 mT applied at $\phi$ = 150º (cyan) and 330º (magenta) for the Hall device of Sample A-w/o with $d$ = 20 μm. $t_{Co}$ and $t_{Ir}$ are 1.3 nm and 0.5 nm, respectively. (b) $\mu_0 H_{sw}$ as a function of $\phi$ for the UD and (c) DU switching events. The black solid circles denote the experimental data while the red curves represent the numerical fit.

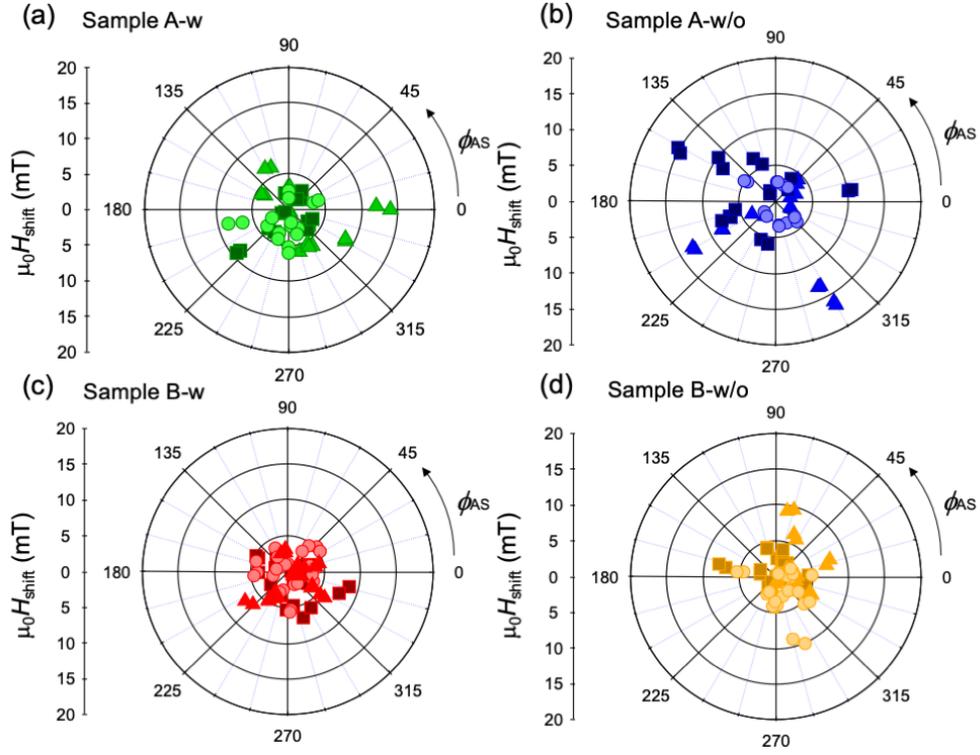

**Figure 6** Pole figures of $\mu_0 H_{shift}$ as a function of $\phi_{AS}$ for the Hall devices of (a) Sample A-w, (b) Sample A-w/o, (c) Sample B-w, and (d) Sample B-w/o, where $t_{Co}$ and $t_{Ir}$ are 1.3 nm and 0.5 nm, respectively. The data points were obtained from 25 devices for Sample A-w, 23 devices for Sample A-w/o, 36 devices for Sample B-w, and 35 devices for Sample B-w/o. Those data points involve both the UD and DU switching events for the Hall devices with $d$ = 5 μm (circles), 10 μm (triangles), and 20 μm (squares).